# *Hubble Space Telescope* Imaging and Spectroscopy of the Sirius-Like Triple Star System HD 217411

J.B. Holberg[1], S.L. Casewell[2], Howard E. Bond[3], M. R. Burleigh[2], and M.A. Barstow[2]

[1] *Lunar and Planetary Laboratory, Sonett Space Science Building, University of Arizona, Tucson, AZ 85721, USA*
[2] *Department of Physics and Astronomy, University of Leicester, University Road, Leicester LE1 7RH, UK*
[3] *Department of Astronomy and Astrophysics, Pennsylvania State University, University Park, PA 16802, USA*

1st September 2011

**ABSTRACT**

We present *Hubble Space Telescope* imaging and spectroscopy of HD 217411, a G3 V star associated with the extreme ultraviolet excess source (EUV 2RE J2300-07.0). This star is revealed to be a triple system with a G 3V primary (HD 217411 A) separated by ~1.1" from a secondary that is in turn composed of an unresolved K0 V star (HD 217411 Ba) and a hot DA white dwarf (HD 217411 Bb). The hot white dwarf dominates the UV flux of the system. However; it is in turn dominated by the K0 V component beyond 3000 Å. A revised distance of 143 pc is estimated for the system. A low level photometric modulation having a period of 0.61 days has also been observed in this system along with a rotational velocity on the order of 60 km s$^{-1}$ in the K0 V star. Together both observations point to a possible wind induced spin up of the K0 V star during the AGB phase of the white dwarf. The nature of all three components is discussed as are constraints on the orbits, system age and evolution.

## 1 INTRODUCTION

Among the pioneering discoveries of the extreme ultraviolet (EUV) all-sky surveys conducted by the *ROSAT* Wide Field Camera (WFC, Pounds et al. 1992) and the *Extreme Ultraviolet Explorer* (*EUVE,* Bowyer et al. 1994) was the detection of EUV excesses associated with apparently isolated field stars containing unsuspected hot white dwarf (WD) companions (Barstow & Holberg 2003). One of these sources was identified with the *V*= 9.7 magnitude, G star HD 217411. It was among approximately 30 other EUV and UV sources that are now known to harbour close and difficult to resolve hot white dwarf binary components (Holberg et al. 2013). For HD 217411 the EUV excess source is designated 2RE J2300-07.0, while the WD associated with the excess is listed as WD 2257-073 in the McCook & Sion (1999).

Systems such as HD 217411 are of interest on several levels. First, as EUV and UV sources, the hot WD is the descendant of a main sequence star that has relatively recently evolved to a degenerate remnant. Moreover, the initial mass of the WD must have been greater than the present main sequence component(s). This establishes the potential criteria for studying an isolated stellar system that can be employed to define the initial-final mass relation (IFMR), which relates the initial mass of the WD progenitor to the final mass of WD. Traditionally the empirical IFMR has been determined using globular and open clusters and their observable ensembles of WDs (i.e. Dobbie et al. 2009; Casewell et al. 2009). These methods ultimately depend on such parameters as cluster distances and turn-off ages. The availability of WDs in non-cluster binary or multiple stars systems for the most part depends on a different set of parameters such as stellar main sequence masses and stellar evolutionary tracks and thus complement cluster methods. If the WD plus main sequence systems can be resolved then the stage is potentially set for very direct techniques such as astrometric masses to define, or constrain, the evolutionary state of the system. Current examples include Sirius A/B (Liebert et al. 2005) and Procyon A/B (Liebert et al.

2013). Additionally, estimates of dynamical masses are critical to the establishment of parallel empirical tests of the degenerate mass-radius relations (MRR) that are independent of simple spectroscopic mass estimates (see Holberg et al. 2012) and that can be used to compare with theoretical MRR.

HD 217411 (BD -7° 5906) is a member of a broad class of main sequence or post main sequence stars orbited by white dwarfs. These systems have been designated 'Sirius-Like Systems' (SLSs) and are cataloged in Holberg et al. (2013). They include both resolved systems, susceptible to study from the ground or with *HST*, and unresolved systems where space-based spectroscopic or astrometric data is needed to define the system. Interestingly, HD 217411 now qualifies in both these respects. The system is partially resolved, from the ground and from space, but as we show its essential nature is only apparent from space-based observations including spectroscopy.

Prior to the discovery of its EUV excess HD 217411 was listed in the Washington Double Star Catalog (WDS J23006-0704AB), being first discovered in 1938 by Rossiter (1943) as a close double (RST4712) with a 1.0 arc second separation and visual magnitude estimates of 9.82 and 11.0. However, this companion was never associated with the EUV excess.

## 2 OBSERVATIONS

### 2.1 Imaging

HD 217411 was among the 18 EUV/UV excess SLS targets that were initially part of a successful *Hubble Space Telescope* snapshot programme (Barstow et al. 2001) that attempted to resolve the WD components in these systems. Eight out of 17 of these targets were successfully resolved using WFPC2 and the remaining unresolved systems were assumed to have separations of less than 0.08″, the resolution limit of the images. Observations of HD 217411 were not reported at the time, as they became available after publication of the snapshot program results. The WFC2 imaging, obtained on 26 June 2001 with the F300W filter centered at 2987 Å, showed the G3 V star, HD 217411 A, and the presumed EUV secondary source (HD 217411 B) located at a separation of 1.115" and at a position angle of 189°. In preparation for the STIS spectroscopy (see below) a second set of WFC3 images was obtained to ensure proper placement of the slit. These images which included the F336W, F275W and F218W filters more clearly resolve the two sources and exhibit a strong wavelength dependence in the relative brightness of the two sources. These three images are shown in Fig. 1.

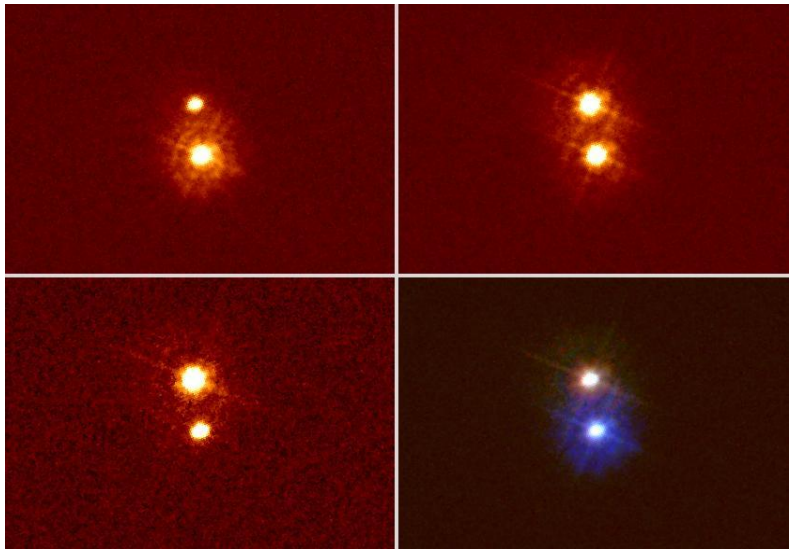

**Figure 1.** HST WFC3 images of the HD 217411 system. Top left: F218W; top right: F275W; bottom left: F336W; bottom right: false-colour rendition of the combined images (blue=F218W, green=F275W, red=F336W). The optical primary star is the northern component of the binary; the primary is the brighter star at 3360 Å, but at 2750 Å the resolved hot companion is nearly as bright as the primary, and at 2180 Å it is 2 mag brighter. Each panel has north at the top and east on the left, and 5.7 arcseconds high.

Subsequent observations of HD 217411 B using the Space Telescope Imaging Spectrograph (STIS) on *HST* were aimed at obtaining spectra of the presumed WD H I Balmer lines of the companion. This was part of a programme (GO 12606, Barstow PI) designed to follow up some of the resolved Barstow et al. (2001) systems to obtain accurate temperature and gravity estimates using the H I Balmer series as well as a gravitational redshift determination for the WD. Surprisingly the STIS spectrometry did not reveal the expected WD spectrum but rather that of an early K star. It was clear from this and the earlier, *IUE*, *FUSE* and COS spectra (Fig. 2), as well as EUV photometry, that a hot WD was obviously present at shorter wavelengths and that HD 217411 B must be a spectral composite source, dominated by a K star in the optical and in the UV by the WD. Hereafter, we designate the G star primary as HD 217411 A, the K star companion as HD 217411 Ba, and the WD 2257-073 as HD 217411 Bb.

**Table 1.** Space-Based observations of WD 2257-073

*HST* Images

| Inst. | Observation | Date | Filter | Exp. Time (s) | Programme | PI |
|---|---|---|---|---|---|---|
| WFPC2 | U5ES210* | 2001 Jun 26 | F300W | 2x12.0 | 8181 | Barstow |
| WFC3 | IBT8050* | 2012 May 26-27 | F218W | 12x15 | 12606 | Barstow |
| WFC3 | IBT8050* | 2012 May 26-27 | F275W | 16x5 | 12606 | Barstow |
| WFC3 | IBT8050* | 2012 May 26-27 | F336W | 4x0.5 | 12606 | Barstow |
| WFC3 | IBT8050* | 2012 May 26-27 | F336W | 12x4 | 12606 | Barstow |

*HST* Spectra

| Inst. | Observation | Date | Grating | Exp. Time (s) | Programme | PI |
|---|---|---|---|---|---|---|
| COS | LB6C06010 | 2010 Sep 30 | G160M | 1652 | 11526 | Green |
| COS | LB6C06020 | 2010 Sep 30 | G130M | 2344 | 11526 | Green |
| STIS | OBT8060* | 2012 Sep 06 | G430L | 3x45 | 12606 | Barstow |
| STIS | OBT8060* | 2012 Sep 06 | G750M | 3x515 | 12606 | Barstow |

*FUSE* Spectra

| Inst. | Observation | Date | Aperture | Exp. Time (s) | Programme | PI |
|---|---|---|---|---|---|---|
| *FUSE* | A0541010000 | 2000 Jun 28 | LWRS | 5082 | A054 | Burleigh |

*IUE* Spectrum

| Inst. | Observation | Date | Aperture | Exp. Time (s) | Programme | PI |
|---|---|---|---|---|---|---|
| *IUE* | SWP46146 | 1992 Nov 06 | LARGE | 2700 | NC119 | Barstow |

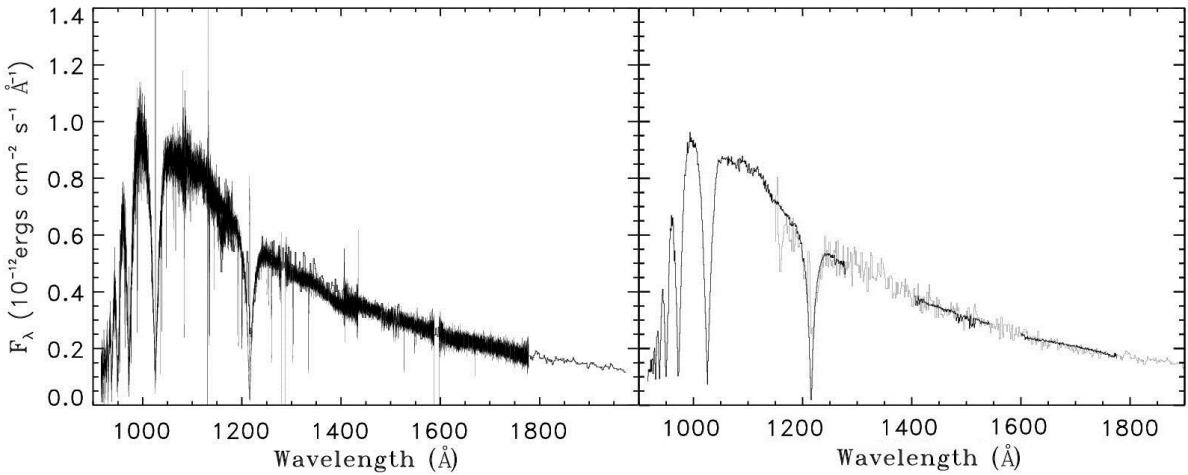

**Figure 2.** Left: The UV spectral energy distribution of WD 2257-073. Shown here are the 900 Å to 1150 Å *FUSE* spectrum, the COS G130M and the G160M spectra, and the *IUE* SWP spectrum (histogram). The gaps in the COS spectra are due to the two detector halves (An electronic colour version of this figure is available.) Right: The merged and re-binned UV spectrum of WD 2257-073 constructed from the data on the left. The 916 Å to 1280 Å data are a combination of the *FUSE* and COS G130L data. The heavy line between 1405 Å and 1777 Å is the re-binned G160L data. The light histogram spectrum is the *IUE* data.

## 2.2 UV Spectroscopy

The WD is most effectively studied spectroscopically at UV wavelengths. A single low dispersion *IUE* SWP spectrum of HD 217411 was obtained in 1992 and a single *FUSE* observation of HD 217411 was obtained in 2000. In addition, there were COS G130M and G160M spectra obtained in 2010. All spectra were obtained from Mikulski Archive for Space Telescopes (MAST) and are also displayed in Fig. 2.

The various UV HD 217411 spectra shown in Fig. 2 (left) cover the entire 900 Å to 1970 Å region with substantial overlap but at differing spectral resolutions. In general the flux levels are all consistent except for the two COS spectra which show a ~10 per cent mismatch beginning at around 1350 Å. It is straight forward to resolve this discrepancy by considering the *IUE* spectrum, which appears to be consistent with all of the other fluxes in Fig. 2. By this criterion the COS G130M spectrum is clearly low, beginning at 1350 Å. We have no explanation for this and disregard the G130M fluxes beyond the spectral gap at 1380 Å in favour of *IUE*. Rather than separately analyze the *FUSE* and COS spectra in detail we construct a single composite spectrum by merging and resampling the spectra into one angstrom bins (Fig. 2, left). In this way it is possible to both consider the Lyman lines and to accurately define the entire UV continuum level with a single data set. Prior to resampling the *FUSE* and COS spectra, the ISM absorption lines and the geocoronal emission lines in the Lyman line cores were masked and interpolated over. This leaves a faithful observational representation of the entire H I Lyman series and UV continuum. The region between G130M spectral gap and the beginning of the G160M data the spectrum is defined only by *IUE* as well as wavelengths beyond 1777 Å. The spectral merging and resampling is archived by averaging the data weighed by their mutual flux uncertainties. The final result is shown in Fig. 2 (right).

## 2.3 STIS Spectrometry (HD 217411 B)

STIS observations of HD 217411 B were obtained on 6 Sept. 2012 using the G430L (R ~ 500) and G750M (R ~ 5000) gratings and the 52x0.2" slit. In Fig. 3 (left) we show both the G430L and G750M spectra along with the optical extension of the UV model flux from WD2253-073 indicating that the composite spectral nature of the K star. In section 3.2 we discuss the template spectrum shown in Fig. 3 (left) that is used to identify the HD 2174111 B as a K 0 V star.

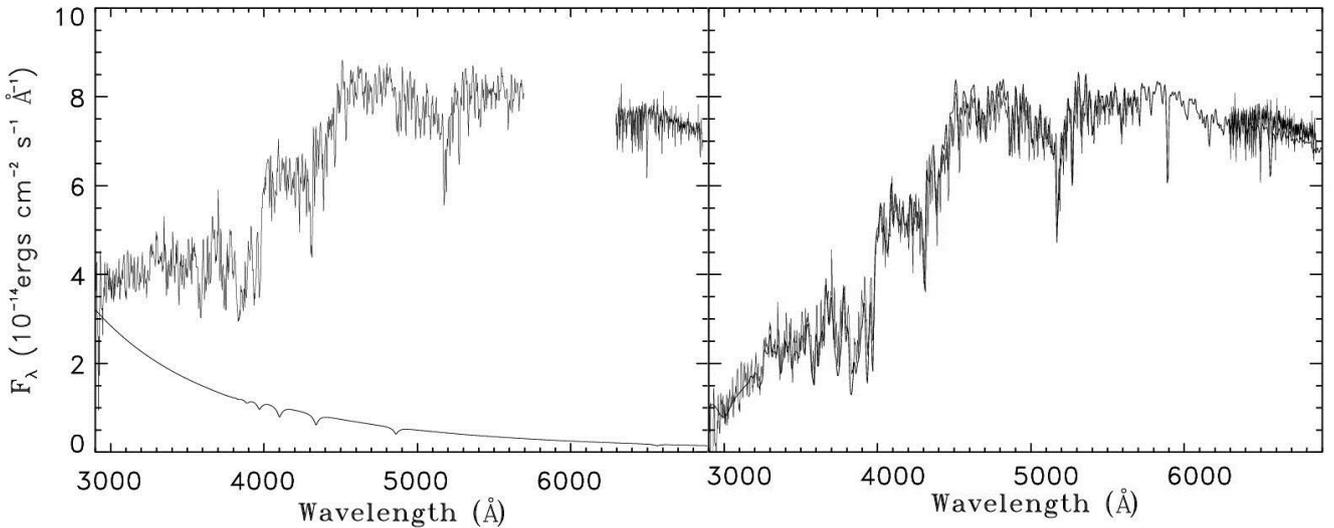

**Figure 3**. Left: The STIS spectra of HD 217411 B and the optical extension (solid curve) of the DA model atmosphere spectrum of WD 2257-073. Right: The WD subtracted STIS spectra of HD 217411b. Over plotted is a normalized Pickles (1998) spectrum for a K0 V star. (An electronic colour version of this figure is available.)

## 2.4 Astrometric Observations

In Table 2 we list the visual and *HST* imaging observations of the separation and the position angle of HD 217411 AB over a 75 year period. Visual separations, position angles and component magnitudes are recorded for five observations at three epochs, 1938, 1943 and 1950 and exhibit orbital motion. These two visual components (A and B), however, are not directly associated with the EUV source 2RE J2300-07.0. As we have shown, HD 217411 B must be an unresolved spectral composite harbouring a close hot WD. Also listed in Table 2 are results from two epochs of *HST* imaging. For the 2012 observations the measurements are the average of the four different filters. There is clear evidence of position angle changes in both the visual data as measured by Rossiter (1950) and between the two *HST* observations separated by 11 years. In Fig. 4 these position angles are plotted as a function of time. The Rossiter visual observations have been assigned a representative uncertainty of 2°. Also shown is a best fit linear representation of apparent the mean motion having a slope of -0.142° yr$^{-1}$. The determination of the slope is obviously dominated by the two *HST* points which have uncertainties of ~0.04°. Over an arc of ~ 11°, the mean motion need not be constant, but a rate of 0.142° yr$^{-1}$ corresponds to an estimated orbital period of 2500 yrs. On the other hand the separation has remained relatively constant at 1.1". At our estimated distance of 143 pc (see Section 3.1) this corresponds to a projected separation of 173 AU. Using the three component masses from Table 3 and Kepler's Third Law yields a substantially lower orbital period of 1500 yrs, implying that physical orbital separation is either significantly larger or that the apparent mean motion during the current epoch is low.

Examinations of the images of the B component in Fig. 1 show no direct evidence of it being resolved. The FWHM of its image profile is essentially equal to that of the A component, 0.07", in the F275W filter in which both the K0 and WD components contribute to the flux. A more stringent test is to look for a shift in the photocenter when going from the F218W filter (dominated by the WD) to the F336W filter (dominated by the K0 component). This test is unfortunately compromised by the asymmetric broadening of the images in the short F336W exposures by the WFC3 "shutter jitter", and by saturated pixels in the images of the A component in the long exposures in this filter. We find an approximate upper limit to the projected separation of about 0.01". This corresponds to a semi-major axis of < 1.6 AU and an orbital period of < 1.7 yrs. This has implications for the expected level of radial velocity variations discussed in section 3.4 as well as the nature of the K0 V component.

**Table 2**
**HD 217411 AB Separations and Position Angles and Magnitudes**

| Date | Separation (") | Position Angle* (°) |
|---|---|---|
| 1938.528 | 1.06 | 198.9 |
| 1938.900 | 0.90 | 196.4 |
| 1943.900 | 0.99 | 199.5 |
| 1950.818 | 1.15 | 191.9 |
| 1950.924 | 0.99 | 193.5 |
| 2001.4853 | 1.115±0.001 | 189.10± 0.04 |
| 2012.4027 | 1.103±0.002 | 187.52± 0.05 |

* Rossiter position angles are as originally published, precession corrections are negligible (< 0.08°). The HST position angles are equinox J2000, not the equinox of the observation date.

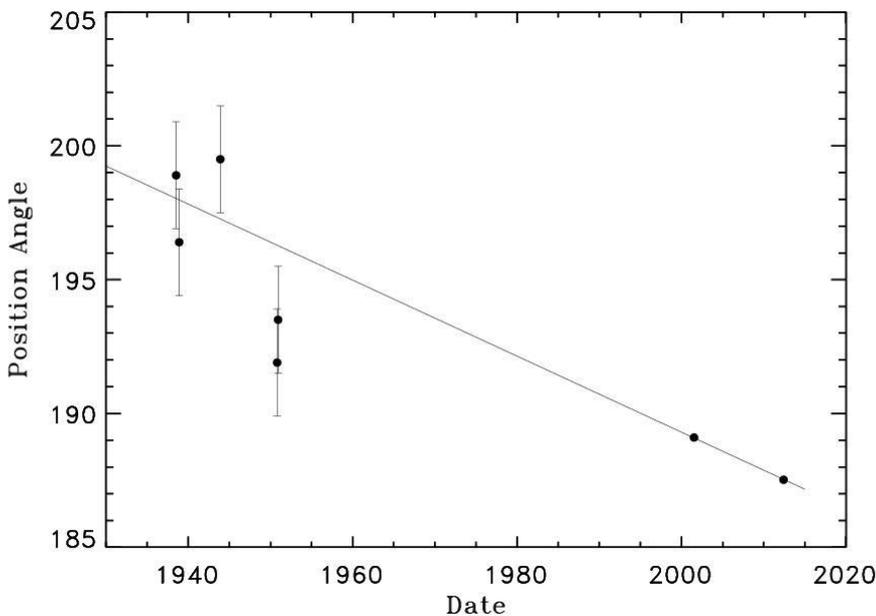

**Figure 4**. The apparent mean motion of the HD 217411 A and B pair. the two very precise points on the right are from the *HST* images. The five points on the left are the visual observations of Rossiter (1950), which were assigned a characteristic uncertainty of 2°. The straight line is a 'best linear fit' to the data; however, the slope is largely determined by the high precision of the *HST* points.

## 3 ANALYSIS

Prior studies of HD 217411, including photometry and spectral type determinations, often refer to the unresolved system in which the A, Ba and Bb components are not distinguished. It is now possible to fully resolve A and B as well as to make useful estimates of individual magnitudes and spectral types for all three components. These estimates are essential to better defining the distance, the nature and orbital state of the system and to constrain its evolutionary history.

### 3.1 HD217411 Bb (2RE J2300-070.0 = WD2253-073)

In Fig. 5 we have fit the Lyman line region of the *FUSE* and COS G130L spectra using a grid of model atmospheres derived from the non-LTE code *TLUSTY* (see Hubeny & Lantz 1995 and references therein). Only pure-H DA models were considered since Barstow et al. (2014) and Lallement et al. (2011) found no photospheric features due to heaver elements in the *FUSE* spectrum, a result borne out by the COS spectra which also show only photospheric H. Fitting included the Lyman series lines between 927 Å and 1050 Å and also included the (COS) Lyman α profile between 1180 Å and 1240 Å. Excluded were ~2 Å segments in the line cores of Lyman β and γ and a 5 Å segment in the Lyman α core to exclude geocoronal emission. Such an arrangement makes optimal use of the Lyman series and deemphasizes the continuum. Our best fit model (shown in Fig. 5) yields $T_{eff}$ = 37200 ± 300 K and log g = 7.80 ± 0.05. The resulting $\chi^2/\nu$ statistic is 0.57, where ν is the degrees of freedom. This indicates that the uncertainty per 1Å bin may be over estimated. This result should be compared with 38010 K and log g = 7.84 (Barstow et al. 2010) and 36680 ± 90 K and log g = 7.62 ± 0.03 (Kawka & Vennes 2010) where only the Lyman lines, Lβ and higher, were fit. If we exclude Lyman α from the fit our result is $T_{eff}$ = 37200 ± 300 K and log g = 7.75 ± 0.03. With respect to Kawka & Vennes (2010), our fit yields a WD that is now more massive (0.58 ± 0.02 $M_\odot$ vs 0.49 ± 0.01 $M_\odot$) and more distant (143 ± 5.6 pc vs 75 pc to 105 pc). In addition to the effective temperature and surface gravity, addition of a modest interstellar H I column density improves the fit in the Lyman α profile. Vennes et al. (1998) made estimates of the interstellar H I column as a function of $T_{eff}$ from the EUV photometry, ranging from $2.0\times10^{19}$ to $5.0\times10^{19}$ cm$^{-2}$. The Lorentz wings of the Lyman α profile are marginally sensitive to H I column densities and we find that columns exceeding $4.0\times10^{19}$ cm$^{-2}$ are ruled out by our data.

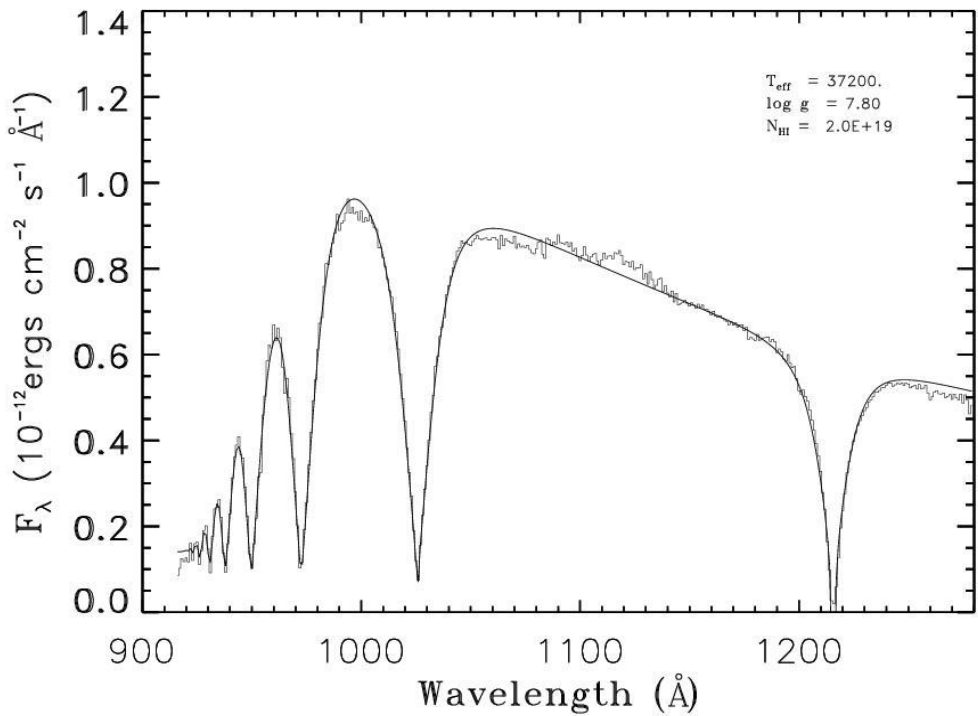

**Figure 5.** The H I Lyman line region of WD. The histogram data is the merged and re-binned fuse and COS G130L spectra. The smooth curve is the best fitting model.

The UV flux levels of the model spectrum yield an expected V mag. of 15.003 for this star. Using the synthetic photometric techniques of Holberg, Bergeron & Gianninas (2008) an estimated distance of 143 ± 5.6 pc is found. This provides the most reliable distance to the system and is in agreement with the distance moduli of the A and B components (see below). The mass and the cooling age of the WD are estimated[1] to be 0.582 ± 0.021 $M_\odot$ and 4.72 ± 0.18 x$10^6$ yrs respectively. Assuming the WD is the product of single star evolution then the progenitor mass can be estimated from the low-mass end of the Initial-Final Mass Relation of Kalirai et al. (2008) to have been 1.9 ± 0.26 $M_\odot$. This corresponds to an A star on the range to A9 V to A4 V/A5 V. The estimated main sequence lifetime for the lower mass A9 V star is on the order of 2.9x$10^9$ yrs, which helps establish an upper age limit for this system. Alternately, if the present Ba-Bb orbit is very close, then the WD could be the product of common envelope evolution, in which case the progenitor mass and spectral type would be uncertain.

### 3.2  HD 217411 Ba

The WD-subtracted STIS spectra of HD 217411 B are shown in Fig. 3. These observed spectra can be nicely matched with that of a K0 V star from the library of Pickles (1998). The Pickles template spectrum is designated no. 32 and identified as 'rK0V' in the Pickles library having a $T_{eff}$ of 5236 K and a [Fe/H] of 0.5. This particular Pickles spectrum was normalized to the STIS fluxes between 5500 Å and 5600 Å using a scale factor of 7.799x$10^{-14}$. The Pickles templates have a spectral resolution of approximately 500 which corresponds to that of the G430L spectrum. As is evident it reasonably represents the STIS data over the entire 3000 Å to 6870 Å wavelength span, with perhaps the exception of a 2.5 per cent to 3 per cent deficit with respect to the G750M fluxes. A similar match is also possible with Pickles' spectrum no. 31 designated 'K0V' having a lower metallicity of [Fe/H] of 0.1. The principal difference with the latter match is that the 'g-band' region between 4000 Å and 4300 Å, shows a pronounced ~10 percent deficit with respect to the G430L fluxes, providing an indication that a higher [Fe/H] is applicable to this star. Thus, we identify HD 217411 as a K0 V star enriched in Fe with respect to the sun. Our Pickles representation of HD 217411 Ba can also be used to estimate the V-band, B-band and *g*-band magnitudes of the K0 V component. In doing this we have used the synthetic photometry techniques of Holberg & Bergeron (2006) using the Cohen and Landolt V-band and B-band filters (Cohen et al. 2003) and the SDSS g-band filters. We estimate V-band magnitudes of 11.76 and 11.71 for B and B + Ba, respectively and 12.4 and 12 for the g-band magnitudes of B and B + Ba, respectively. It should be noted that the *V* magnitude determined here is considerably fainter than the mean of the visual estimates (*V* = 11.02) from Rossiter (1950). Given the relatively young system age determined above, a zero-age main sequence absolute magnitude age for the K0 V star of $M_v$ = 5.9 would seem a reliable estimate. This yields a distance of the B component of 148 pc in good agreement with the distance estimate of the WD determined above. The mass corresponding to the K0 V star is approximately 0.79 $M_\odot$.

The WD-subtracted G750M spectrum of the K0 V star is shown in more detail in Fig. 6 along with the template for the K0.5 V star 40 Eri A (HD 26965). This template, obtained from the *Elodie* archive of R=10000 spectra, has been modified in the following manner. In order to match the G750M spectrum, the template is Doppler shifted to 0.0 km s$^{-1}$ and rotationally broadened and then resample onto the G750M wavelength scale. Several aspects of the K0 V star are evident. First, there is no clear Hα absorption feature. The expected absorption line is presumably filled by a low level of emission. Second, the stronger photospheric absorption features are clearly broadened, presumably by rapid rotation. It is found that the G750M spectrum is best matched by the template spectrum after rotationally broadening it with a *v sin i* = 60 km s$^{-1}$. This is a surprisingly large rotation and implies that the K0 V star has been spun up by wind driven angular momentum transfer (Jefferies & Stevens 1996a) that occurred during the mass loss phase of the present white dwarf. Finally, there is no clear evidence in the G750M spectrum for enrichment by s-process material, for example the Ba II line at 6349.9 Å.

### 3.3  HD 217411 A

The spectral type of the primary component of this triple system has been variously reported as G3V (Houk & Swift 1999), G4 V (Torres et al. 2006), and G4 V/G5 V (Cutispoto et al. 1999). Based primarily on photometry discussed below, we shall adopt a spectral type of G3 V. There exists relatively little independent photometry visual band photometry of HD 217411. A *V* magnitude of 9.7 ± 0.02 is given in Fekel (1997). Alternately, the Tycho BT and VT magnitudes (Wright et al. 2003) can be transformed to a *V* magnitude and *B-V* colour (Mamajek, Meyer & Liebert 2003) to give a consistent estimates of *V* = 9.715 and *B-V* = +0.66. It is useful to have estimates of magnitudes and colours that refer to the uncontaminated G3 V star. Using the observed fluxes of the B component (see Section 3.2) it is possible to estimate uncontaminated optical magnitudes and B-V colour of the A component. We find *V* = 9.85 ± 0.03 and *g* = 9.8 ± 0.03 and *B-V* = +0.64. This estimate of the uncontaminated colour places the star closer to the spectral classification G3 V. Interestingly the *V* magnitude estimated here is close to the mean value (*V* = 9.81) of the Rossiter observations. Thus, we find HD 217411 A to be slightly fainter and of somewhat earlier spectral type than the primary of EUV 2RE J2300-070 considered by Vennes et al. (1998), who estimated a distance of 72 – 105 pc.  In Sections 3.1 and 3.2 photometric distances of the Ba and Bb components yield respective distances of 148 pc and 143 pc, and a corresponding distance modulus of 5.8. At this distance the absolute magnitude of the primary is $M_v$ = 4.05, indicates that the G star may be of earlier spectral type and/or perhaps over luminous.   HD 217411 is also observed to exhibit a moderate amount of chromospheric activity (Mullis & Bopp, 1994) and to have a *v sin i* of 3.4 km s$^{-1}$ (Fekel 1997).

---

[1] See http://www.astro.umontreal.ca/~bergeron/CoolingModels.

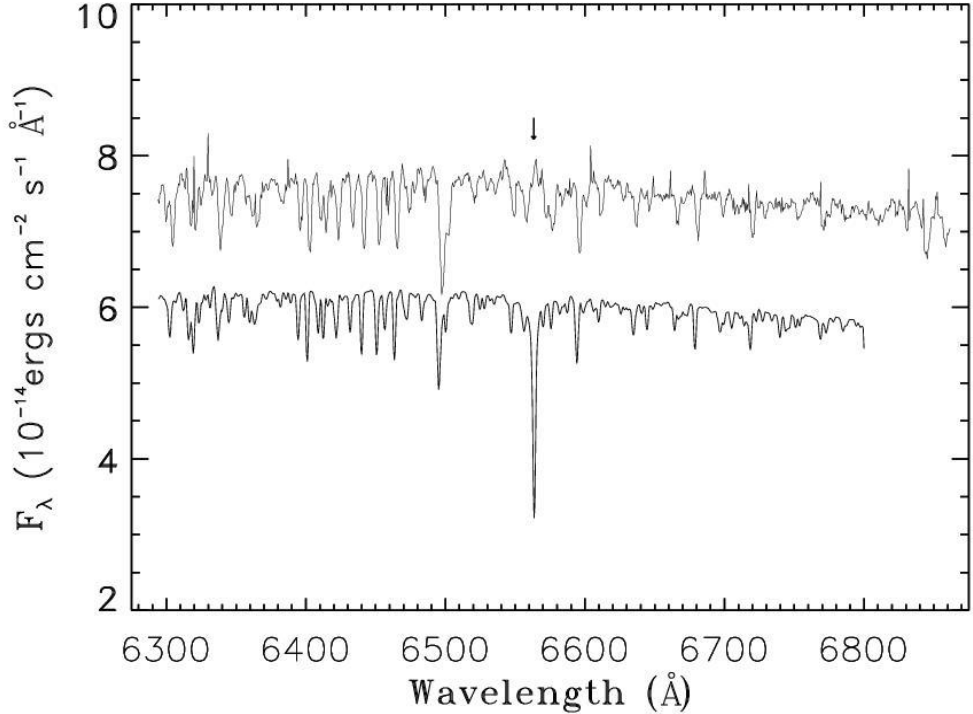

**Figure 6.** A comparison of the G750M spectrum (shifted to an 'air' wavelength scale) for the K0 V star (top) and a template K0.5 V star (bottom).  In this figure, the flux of the template star, 40 Eri A, has been reduced by a factor of 100 and the wavelengths have been doppler shifted to a rest frame of 0.0 km s$^{-1}$ and spectrum rotationally broadened with a  *v sin i* = 60 km s$^{-1}$. The arrow indicates the expected location of Hα, which in contrast to the template, shows a possible slight emission, in place of a strong absorption feature.

**3.4 Radial Velocities**

There exists a limited amount of radial velocity information on the components of the HD 217411 system.   Vennes et al. (1998) found no evidence of radial velocity variations in the G star primary at the 0.8 km s$^{-1}$ level over 400 days and measured a system γ velocity of +22.4 ± 0.8 km s$^{-1}$ during the epoch 1996; this is consistent with Torres et al. (2006) who also find a radial velocity of +23.0 ± 1.3 km s$^{-1}$ in the 1996 epoch.  The lack of radial velocity variations in the primary is not surprising since the WD is associated with a distant K0 V component having a large separation and ~ 2500 yr. orbit. Employing the results of Holberg et al. (2013) the estimated change in the radial velocity this system is expected to be < 5.4 m s$^{-1}$ per year.  On the other hand, significant variations in radial velocity can be expected for the Ba – Bb subsystem.

The possibility of radial velocity variations in the K star WD subsystem are of considerable interest.  With respect to the Bb (WD 2273-073), Lallement et al. (2011) provide an extensive discussion of the various velocity components present in the *FUSE* and COS spectra. Although Lallement et al. find no credible photospheric features (other than the Lyman lines) in the spectrum of WD 2257-073 they did estimate a photospheric velocity by cross-correlating synthetic Lyman profiles (β, γ and ε) with the *FUSE* spectra obtaining a heliocentric velocity of -7.5 (+8.4, -7.5) km s$^{-1}$.   If the heliocentric velocity of the Lyman lines is correct then subtracting the +22.4 km s$^{-1}$ system velocity would give an apparent orbital velocity of -29.9 ± 9 km s$^{-1}$ for the WD for the epoch 2000.491.   For reference the expected gravitational redshift for the WD is +24.7 km s$^{-1}$, implying an actual orbital velocity of ~ -55 km s$^{-1}$ at that time.

Our G750M spectrum, originally intended to measure the gravitational redshift of the WD component, can be used to measure the radial velocity of the K0 V component by converting the STIS wavelength scale to 'air' and cross-correlating the data with the template from the ELODIE archive[2] in Fig. 6.  Using cross-correlation windows at 6370Å to 6520Å and 6580Å to 6700Å give observed velocity estimates of +23.4 km s$^{-1}$ and +18.5 km s$^{-1}$ from which we estimate and observed velocity of = +20.9 km s$^{-1}$.  Subtracting the system $\gamma$ velocity of +22.4 km s$^{-1}$ gives an orbital velocity of -1.5 km s$^{-1}$ for the K0 V star with respect to the center of mass for the epoch 2012.685.  Assuming respective upper limits of 1.7 yrs and 1.6 AU for the orbital period and separation of the Ba-Bb pair from section 2.4 gives a lower limit to the expected orbital amplitude velocity of 28 km s$^{-1}$.  Employing the results of Holberg et al. (2013) a corresponding lower limit to the radial velocity acceleration of the K0 V star is on the order of 28 km s$^{-1}$ yr$^{-1}$.  Thus over the period of less than a year velocity changes could easily be detected in this star, if it can be sufficiently resolved with respect to component A.

The G570M spectrum in Fig. 6 shows two important aspects.  First, there is no H$\alpha$ absorption feature, if anything there is a slight emission at the expected location.  Second, the remaining photospheric features are significantly broadened with respect to the initial 40 Eri B template.  Assuming that this is due to rotational broadening a *v sin i* of 60 ± 10 km s$^{-1}$ can be estimated.  This has been applied to the template spectrum in Fig. 6 and indicates a rather rapid rotation for the K0 V star.

**Table 3**
**HD 217411 Components**

| Parameter | HD 217411 A | HD 217411 B | WD 2253-073 |
|---|---|---|---|
| Spectral Type | G3 V | K0 V | DA1.4 |
| *V* mag. | 9.85[1] | 11.75[1] | 15.003(0.03)[3] |
| *B-V* | +0.64[1] | +0.85[1] | -0.27[1] |
| *g* mag. | 10.06[1] | 12.10[1] | 14.72[3] |
| $T_{eff}$ (K) | 5670[2] | 5236[2] | 37200(300)[3] |
| Mass ($M_\odot$) | 0.96 | 0.79 | 0.58(0.02)[3] |
| Age (Myr) | …. | ….. | 4.73(0.18)[3] |

[1]Synthetic Photometry, [2] Pickles (1998),[3] this paper

## 3.5 Photometric Variability

In addition to the space-based photometric and spectroscopic data discussed here there also exists a low level modulation of the total light from the system.  HD 217411 is among a number of Sirius-Like systems that have been photometrically monitored by the WASP (Wide Angle Search for Planets) Survey (Pollacco et al. 2006).  Several of these unresolved systems, including HD 217411, show low levels of photometric variability (Faedi et al. 2014).  In particular HD 217411 exhibits a sinusoidal light curve with a period of 0.61091 days (14.662 ± 0.001 hrs) with a peak-to-peak amplitude of 2 per cent.  The WASP data were obtained over periods of months during observing seasons in 2008, 2009, and 2010 in broad-band white light with large 14" pixels.  Thus the variations refer to the total light of the system and appear to be coherent over the multi-year observing period, with no evidence of eclipses.

The precise source of this modulation remains unidentified.  A rotational modulation of the G3 V star (which provides 85 per cent of the total light) is unlikely, since the relatively short period would imply a large *v sin i* (~ 80 km s$^{-1}$), while the measured *v sin i* (Fekel 1997) is only 4.3 km s$^{-1}$.  Likewise, the WD is not a likely to be a direct source as it provides only 0.7 per cent of the visible light.  The contribution of the K0 V star on the other hand is on the order of 15 per cent of the total light, therefore a 2 per cent modulation of the total light would require an intrinsic modulation of 13 percent peak-to-peak in the K 0 star, a rather large amount.

## 4 DISCUSSION

The HD 217411 system is found to be a somewhat younger and more compact analog of the well known Sirius-Like triple system 40 Eri A/B/C, which consists of a WD (B) and M5 V (C) star pair orbiting a distant (415 AU) K0.5 V primary (A) with an orbital period of ~ 8000 yr (Heintz 1974).  The WD and M star pair has a separation of 35 AU with a mutual 250 yr orbital period.  In contrast the corresponding components of HD 217411 are all more massive and of earlier spectral type (DA1.4/K0 V/G3 V), while the system itself is approximately three to four times more spatially compact. In spite of the fact the B-Ba components remain unresolved, UV spectra of the DA star provide key information necessary to better define the two main sequence components and to constrain the most likely stellar evolution of the system. For example the subtraction of the WD flux permits a firm identification of HD 217411 B as a K0 V star. This in turn allows a refinement of the photometry and spectral type of the primary HD 217411 A. The estimated system distance of 143 pc is also established by the DA star and is consistent with what is known about main sequence components. The clear identification of the K0 star in turn allows a better definition of the photometry and spectra type of the primary HD 217411 A. From the mass of the DA star ($0.58 \pm 0.02$ $M_\odot$) estimates of the progenitor mass provide a system age of less than 2.9 Gyr.

---
[2] http://atlas.obs-hp.fr/elodie/

The principal uncertainty remains the orbital separation of the WD and K0 V subsystem. There is little evidence that the two stars are in extreme proximity or that they are currently undergoing active mass transfer. The mass of the WD (0.58 $M_\odot$) is very close to the mean of the single star WD mass distribution for C/O core degenerates. Further, the K0 V star shows no evidence of prominent emission which might arise from irradiation by the WD that is often seen in post Common Envelope binaries. There are, however, two lines of evidence that suggest that the WD and K0 V star may have interacted in the past. First, is the rather large $v \sin i$ indicating that the K0 star was spun up through wind induce angular momentum transfer during the AGB phase of the WD. Second, is the 0.61 day photometric variability. If this variability is due to the K0 V star, then it implies a rotational velocity on the order of 70 km s$^{-1}$ which is certainly consistent with our estimated $v \sin i$ of $60 \pm 10$ km s$^{-1}$. In both respects this is a situation that resembles another EUV Sirius-Like system 2RE J0357+283, where a hot DA star orbits a rapidly rotating ($v \sin i = 141 \pm 5$ km s$^{-1}$) K2 V star (Jeffries, Burleigh, & Robb 1996). The proposed mechanism for the rapid rotation is wind-accretion induced rapid rotation of the K-dwarf occurring from AGB mass loss from the WD progenitor; the so called WIRRing stars (Jeffries & Stevens 1996b). Interestingly 2RE J0357+283 also exhibits a photometric variability of 0.15 magnitudes and a pronounced variability in the Hα emission. All of this suggests at the WD K0 V system in HD 217411 may be a similar binary where the observed photometric variation is due to a star spot.

**ACKNOWLEDGEMENTS**

J.B.H. acknowledges support from Space Telescope Science Institute grant HST-GO-120606.01-A and NASA Astrophysics Data Programme grant NNX1OAD76 and NSF grant AST-1008845. MAB and MBR are supported by STFC, UK. SLC acknowledges support from the College of Science and Engineering at the University of Leicester. We also wish to thank Saurav Dhital for help with ELODIE templates. This research has made use of the *WD Catalog* maintained at Villanova University, and the *SIMBAD* database, operated at CDS, Strasbourg, France. This research has made use of the Washington Double Star Catalog maintained at the U.S. Naval Observatory. Some of the data presented in this paper were obtained from the Mikulski Archive for Space Telescopes (MAST). STScI is operated by the Association of Universities for Research in Astronomy, Inc., under NASA contract NAS5-26555. Support for MAST for non-HST data is provided by the NASA Office of Space Science via grant NNX13AC07G and by other grants and contracts.